# MULTISCALE MODELLING TO ESTIMATE SPALL PARAMETERS IN METALLIC SINGLE CRYSTALS

S. Madhavan[*1], V.R. Ikkuthi[1], P.V. Laxminarayana[2] , M Warrier[1]

[1]Bhabha Atomic Research Centre, Visakhapatnam, Andhra Pradesh, 530012, India

[2]Dept. of Nuclear Physics, Andhra University, Visakhapatnam, Andhra Pradesh, 530003, India

[*]Corresponding author e-mail: madhavan73@gmail.com

## ABSTRACT

Modeling dynamics fracture in materials involves usage of hydrodynamic codes which solve basic conservation laws of mass, energy and momentum in space and time. This requires appropriate models to handle elastic-plastic deformation, equation of state, material strength, and fracture. Nucleation and Growth (NAG) damage model is a micro-physical model which computes amount of damage in the material by accounting for phenomena like nucleation, growth and coalescence of voids or cracks. The NAG model involves several material model parameters, such as nucleation threshold, growth threshold, etc. Traditionally these parameters are fitted to experimental void volume distributions. In the present paper we fit these parameters to molecular dynamics (MD) simulations of void nucleation and growth and use the fitted parameters in hydrodynamic simulations in a multi-scale computational approach. Cubic metallic single crystals are subjected to isotropic deformation and the nucleation of voids and their growth were post-processed from the simulations. These results are used in an in-house Particle Swarm Optimization (PSO) code to obtain NAG parameters for materials of our interest. Using these parameters in a 1D hydrodynamic code developed in-house, fracture parameters such as spall strength and thickness are obtained. The results are validated with published experimental data for Mo, Nb and Cu which have been simulated using the multi-scale model. This paper describes the application of the multi-scale model to obtain the NAG fracture model parameters of Al and its spall data. The results are compared with published experimental results in single crystal Al.

**Keywords:** Dynamic Fracture, Multi Scale Modeling, Single Crystal, Molecular Dynamics, Hydrodynamics, Shockwave

## 1.0 INTRODUCTION

Rapid loading of energy, on a material, produces shock-wave [1-3]. Shockwave loading can be achieved by impact of a high velocity metallic plate or projectile on a target material. High velocities can be achieved using the energies of Explosives, Electromagnetic & Laser systems. Velocities in the range of 10 m/s to 5 km/s can be obtained using these methods [1-6].  As shown in figure 1, upon impact of  the moving material (velocity U) on a target, the shock wave is produced at the region of impact and travels inwards through the impactor with a velocity Us1 and along the target with a velocity Us2, as stress waves. This leads to high strain rates in the rage of $10^4$ to $10^6$ s$^{-1}$, as observed in plate impact experiments [7-10]. High power pulsed lasers produce much higher strain rates beyond $10^6$ s$^{-1}$ [11]. Shock wave produced by such an impact, travels through the thickness of both, the impactor and the target and reaches their free surfaces. Upon reaching free surfaces, relaxing phenomena called rarefaction wave travels in the opposite direction of shock wave and towards inner region of the impactor and target with a velocity Ur1 and Ur2. These two rarefaction waves travel towards each other and eventually interact inside the material, causing a sudden fall of local pressure leading to a tensile stress in the material. The tensile stress can create void nucleation, growth and coalescence in the material leading to fracture. This type of failure is called 'spallation' [12]. The region where spallation occurs depends upon the thickness or the distance traveled by the shock wave in the impactor and the target as well as the density of both. If the thickness of the impactor is few times smaller than that of the target then 'spallation' occurs  close to the other end of impacted surface. If the impact is normal and uniform along surfaces of contact, 1 dimensional shock wave is produced. Thus the spallation is expected to occur along the single  plane on the material as shown in figure 1.

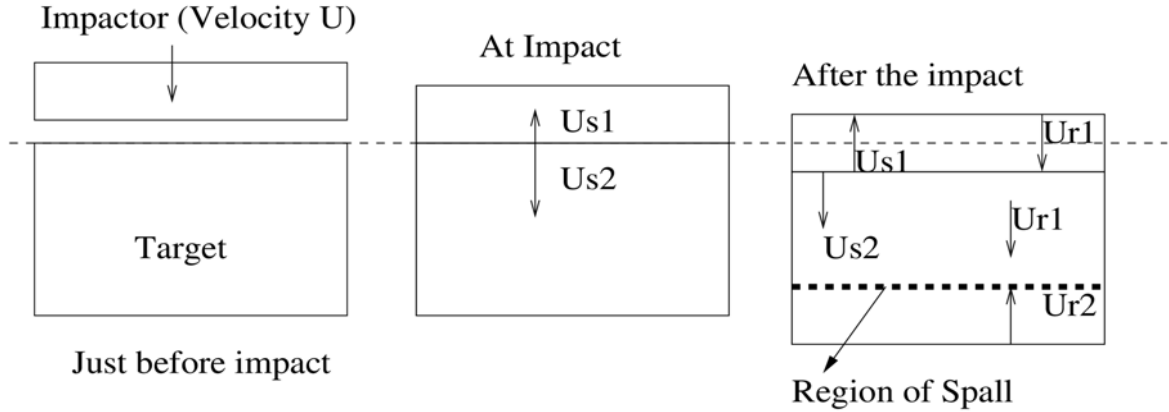

*Fig 1: Schematic of Impact system showing the compressive shock and reflected rarefaction waves*

Hydrodynamic mathematical tools are useful in predicting spallation failure of materials. This in turn requires material parameters such as equation of state, dynamic strength and an appropriate fracture model. Experiments and modeling can be used to obtain these parameters. We have devised a multi-scale modeling for spallation of single crystalline metals. First, atomistic simulations using the Large-scale Atomic Molecular massively Parallel Simulator (LAMMPS) [13] are carried out to simulate isotropic expansion leading to tensile deformation, creation of voids and their growth. The temporal and spatial evolution of the atoms under such a loading gives the ambient pressure and void volume. The Nucleation and Growth (NAG) model [14, 15] relates the instantaneous pressure and void volume of the system. NAG model has 6 adjustable parameters whose details are given in section 2. Using an appropriate optimization, one can fit the unknown parameters of the NAG model from the results of the atomistic simulations. The parameters so obtained are used in the macro scale hydrodynamic models [16-18] to predict the integrity or failure of the materials subjected to impact loading.

It has been shown [18,19] that such a multi-scale methodology works in general. The spall threshold and spall thickness, obtained from hydrocode simulations, form optimized NAG model parameters using MD simulations, agree well with experimental values in case of single crystal metals viz Molybdenum (Mo), Niobium (Nb), and Copper (Cu) as reported [20]. Taking cue from this work, we have developed a Particle Swarm Optimization (PSO) method to obtain NAG parameters from MD simulations. In the present work, we have obtained the spall strength and spall thickness for aluminum (Al) single crystal. These are found to be in good agreement with the published experimental results [21]. The computational methods are described in section 2. The results are presented and discussed in section 3. Finally the conclusion is presented in section 4.

## 2.0 COMPUTATIONAL METHODS

The computational methods used in this work are (i) MD at the atomistic scales using LAMMPS, (ii) PSO to post-process the atomistic simulations and fit the NAG model parameters and (iii) One dimensional hydrodynamic code to model an impact experiment at the macroscopic scales using the NAG parameters obtained from PSO. Schematic of the sequence of execution of the multi-scale model is given in figure 2. The details of the various computational models used in the multi-scale modelling are explained in the following sections (2.1 to 2.4).

### 2.1 Molecular Dynamics simulation of tri-axial deformation :

We have used LAMMPS to simulate the isotropic, tri-axial deformation of a 100 x 100 x 100 unit cell single crystal Mo, Nb, Cu and Al with Embedded Atom Method (EAM) potentials [22-24]. The potentials were validated by obtaining the lattice parameter, cohesive energy, vacancy formation energy and bulk modulus of these elements and comparing them with experiments. Single crystals used in experiments of spall fracture, consists of defects and impurities. In order to account for such defects an edge dislocation is created by removing a half-plane of the crystal and by equilibrating the sample using an NPT ensemble.

A constant time-step of 1femto sec (fs) is used with periodic boundary conditions along X, Y and Z. NPT simulations are performed at 0 bar and 300 K to first equilibrate the system. After equilibration of the system, the barostat was turned off and a strain rate $10^9$ s$^{-1}$ triaxial expansion is applied at constant temperature by changing the box size as $L(t) = Lo(1 + \dot{\varepsilon} t)$, where $Lo$ is the initial box size, $\dot{\varepsilon}$ is the strain rate and $t$ is the elapsed time. This change in box size is carried out every 100 time steps. Therefore the atoms have around 0.1 pico sec (ps) to relax to the increased strain before further deformation. We have varied the time interval between these deformations from 1 to 100 time steps and seen that the pressure time history follows the same curve.

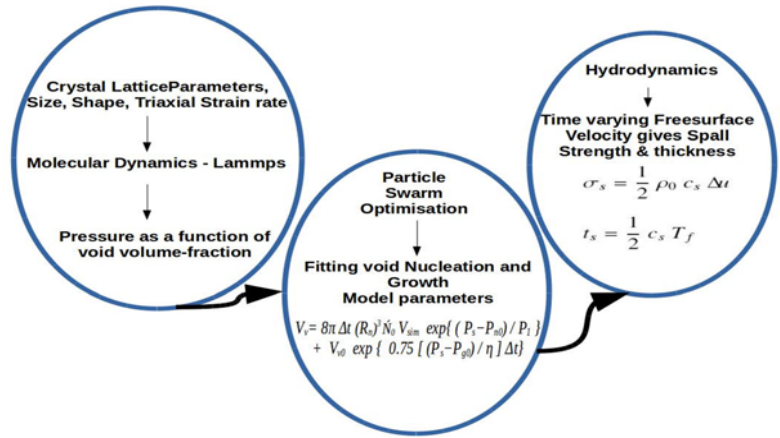

*Fig 2: Flowchart showing the flow of Multi-scale Modelling*

## 2.2 Void Volume Calculation

The temporal evolution of the growth of void was obtained as a function of ambient pressure from the MD simulations as described in [25]. The volume being simulated is divided into voxels (3D equivalent of pixels). Initially each voxel is assumed to be empty. From the positions of the atoms each voxel is designated as empty or filled. The sum total of the empty voxels gives the void volume. In order to study the void size distributions, if any two voxel lie beside each other in any of the three dimensions, they are treated as connected. This goes on recursively and the voxel size gets accumulated over time. In this way the time varying number of independent voids, the total void volume and void size distributions are calculated. The parameters of the NAG model are then fitted using the PSO technique described below.

## 2.3 Particle Swarm Optimization (PSO):

PSO belongs to the general class of evolutionary algorithms which are suitable to solve multi-parameter minimization problems. This method resembles a group of flying birds, hence named so. In order to optimizing a function consisting of 'n' independent parameters, several particles are initiated with an initial random guess for each of the independent variables defined within a reasonable range. During each iteration, the objective function is evaluated for each of the particles and the "best particle" is identified for each iteration and the "globally best particle" over all the iterations is identified. The positions of the particles in 'n' dimensional space are updated in every iteration, so as to get a better solution if it exists. Let the $i$th particle during the current iteration be represented as $X_i = (x_{i1}, x_{i2}, .., x_{in})$ for a "n" parameter problem. If its previous best position is represented as $P_i = (p_{i1}, p_{i2}, .., p_{in})$ and the global best position amongst all particles is represented as $P_i = (p_{g1}, p_{g2}, .., p_{gn})$, then the new incremental value of each independent variable, required to move closer to the solution is :

$$v_{id} = v_{id} + c_1 * r_1 * (p_{id} - x_{id}) + c_2 * r_2 * (p_{gd} - x_{id}) \quad - (1a)$$

and the new position of each particle is updated as

$$x_{id} = x_{id} + v_{id} \quad - (1b)$$

where $c_1$ and $c_2$ constants, $r_1$ & $r_2$ are uniform random numbers between 0 to 1. The second term in the R.H.S of (Eqn. 1a) is called the 'cognition' part that drives the particle to its best position and the third term on the RHS drives the particle to the globally best position [26].

The computational Fortran code for this model was developed by us and has been validated with standard functions used to test global optimization codes, viz. (a) Spherical functions, (b) Rastrigin function, (c) Ackley function and (d) Griewank functions [27]. We have validated the convergence of these functions using our PSO code, up to 10 variables function, with 50 particles. PSO could reproduce the theoretical result accurately.

## 2.4 One Dimensional Hydrodynamic Model :

The model solves the fundamental conservation equations of mass, momentum and Energy in time and spatial co-ordinates. The 1-D model developed in-house follows the formalism of KO [16]. It is formulated using Finite difference method in Lagrangian frame where the hydrodynamic flow of the object is attached to the computational grid. This makes the mass of the discretized units (gridded computational domain) to remain the same throughout the evolution. For each computational grid, the instantaneous local density for any change in volume is calculated at each step of time evolution. The instantaneous local pressure and temperature are obtained using equation of state (EOS) in terms of density and temperature. For materials with mechanical strength, appropriate strength models viz. Johnson-Cook, Steinberg-Guinan, Zerilli-Armstrong are available. The tool supports plane and symmetric geometries (cylindrical and spherical). The model has been extensively validated with available experimental results [16].

High velocity impact of an impactor (or a flyer) on a target produces shock-waves. Shockwave is characterized by its non-linear property and spatial discontinuity of hydrodynamic parameters viz. pressure, density etc. A discontinuous solution of a differential equation ie. shockwave is inexpressible in discretized computational grids. But the inclusion of artificial viscosity helps to smooth out the shock around the discontinuity region so that spatial discretization is handled appropriately [16]. Our 1D hydrodynamic model has this feature to handle shock-wave.

For this work, we have two zones namely the impactor and target. Each zone is assigned with its computational grid co-ordinate locations, values of density, pressure, energy, temperature . The impactor zone grids were given unidirectional velocity pointing towards target. All these are initial conditions. The two ends of the zones are made freely moving boundaries under the influence of hydro-dynamically evolving thermodynamic parameters. The simulation starts with the moment of impact where the two zones are kept in contact by assigning the same co-ordinate values to the respective contact boundaries. 1D Hydrodynamics give temporal variation of free surface (the far end of the target) velocity. From the nature of this curve, the spall strength and spall thickness can be obtained as described in [14, 15] spall parameters like spall strength $\sigma_s$ and spall thickness $t_s$ have been computed using the formula $\sigma_s = 0.5\ \rho_0\ C_s\ \Delta u$ and $t_s = 0.5\ C_s\ T_f$, where $\rho_0$ is the normal density of the target material, $C_s$ is the bulk sound speed in that material, $\Delta u$ is the pull-back velocity in the free-surface velocity profile and $T_f$ is the period of oscillations of the spall signal.

**2.5 Nucleation And Growth (NAG) Fracture Model :**

This model was developed at Standfort Research Institute [14]. According to this model, fracture occurs due to nucleation and growth of voids in ductile materials. Either when tensile pressure in the solid material $P_s$, exceeds the nucleation threshold, $P_{n0}$ of the material or if the tensile pressure exceeds the void growth threshold pressure $P_{g0}$, the voids grow. Thus the total void volume due to both of these is given by

$$V_v = 8\pi \, \Delta t \, (R_n)^3 \, \acute{N}_0 \, V_{sim} \, exp\{ \, ( \, P_s - P_{n0} ) \, / \, P_1 \, \} + \, V_{v0} \, exp \, \{ \, 0.75 \, [ \, (P_s - P_{g0}) \, / \, \eta \, ] \, \Delta t \} \quad - (2)$$

where $\acute{N}_0$ is the nucleation rate of voids ($m^{-3}s^{-1}$) and $P_1$ the pressure sensitivity for nucleation, Both of these are and are material constants. $V_{vo}$ is the initial void volume at each time step and $V_{sim}$ is the volume of the material under tensile pressure. $\eta$ is the material viscosity. The nucleation size parameter $R_n$ is taken to be the size of a unit cell in single crystal. $\Delta t$ is the time-step used in hydrodynamic model. Note that $\acute{N}_0, P_1, P_{n0}, P_{g0} \, and \, \eta$ are unknown parameters which we hope to determine using the results from the atomistic simulations.

**3.0 RESULTS AND DISCUSSION**

**3.1 Results from MD simulations**

The resulting pressure as a function of time during triaxial deformation of single crystal aluminium at room temperature for a strain rate of $10^9$ /sec is shown in figure 3. The magnitude of pressure increases continuously up to 7 GPa after which it turns around and decreases with further expansion. The turn-around of pressure is the result of stress relaxation due to nucleation and growth of voids. The pressure-time profile can be divided in to three regions as described in [18, 28].

1. Void nucleation region AB (fig. 3) : This region corresponds to the nucleation of voids. In this region, voids are generated when the tensile pressure crosses threshold limit. When a void nucleates at "A" there is a local relaxation of the bonds around the nucleated site. The pressure in Fig.3 is the total pressure of the system. The local relaxation at "A" takes some time (~ 1 ps in this case, which is typically the vibration period of an atom) to reflect in the total pressure of the system. At "B" the information is reflected in the total pressure of the system. Moreover, either the void has grown or new voids have formed which further decreases the stresses in the system and reduces the tensile pressure.

2. Growth region BC (fig. 3) : This region corresponds to growth of voids. Void volume generation due to growth of voids occurs and the tensile pressure falls below the threshold for void nucleation. When this happens void nucleation stops and only growth takes place. The threshold for void growth in this region is always lower than the tensile pressure in the system, leading to continuous growth of the voids.

3. Coalescence region (Region beyond C, fig. 3) : This region corresponds to coalescence of voids. In this region, the void volume due to coalescence of the voids, dominates over growth of voids.

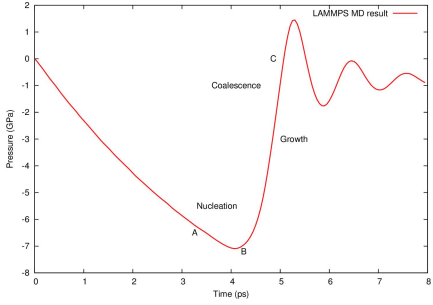

Fig 3: Temporal variation of pressure of 'Al' subjected to triaxial tensile load.

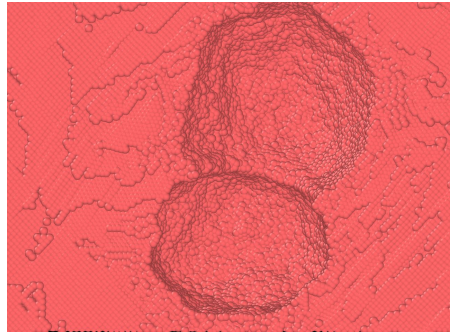

Fig 4: Snapshot of a nucleated void, in a single crystal subjected to tri-axial deformation.

Increase in void volume as a function of pressure for Mo and Al are given in figures 5 and 6 respectively. NAG model establishes the relation between the void volume fraction and pressure. The model has five coefficients which have to be determined from the atomistic simulations as described in section-2.5. These coefficients are determined using the PSO algorithm described in section-2.3. The NAG parameter fitting is also shown in Figs. 5 & 6 for comparison. The relative errors in the fitting are 4 % and 8 % respectively, for Mo and Al. It may be noted that we have used only the nucleation and growth part of the atomistic simulations for the fit, ignoring the coalescence. These errors are better than the earlier fits [19].

Table 1 : Best fit NAG parameters for single crystal Mo & Al, tri-axially deformed at 10$^9$ s$^{-1}$ strain rate

| Sample | $\acute{N_0}$ $(m^{-3}s^{-1})$ | $P_{n0}(GPa)$ | $P_1(GPa)$ | $P_{g0}$ $(GPa)$ | $\eta$ $(10^2\,Pa\text{-}s)$ |
|--------|------------------------------|---------------|------------|------------------|------------------------------|
| Mo | 26.42 x10$^{35}$ | -17.5 | -1.9 | -10.0 | 0.896 |
| Al | 13.06 x10$^{35}$ | -10.11 | -3.68 | -0.93 | 0.7 |

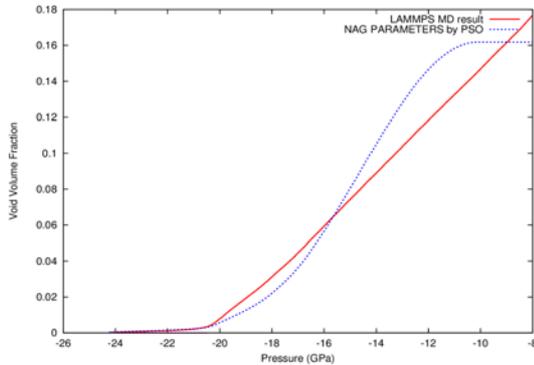

Fig 5: Void Volume fraction obtained from Pressure from LAMMPS & fitted NAG model using PSO method for Mo single crystal.

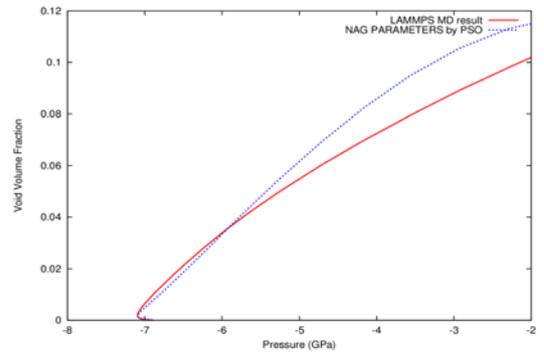

Fig 6: Void Volume fraction obtained from Pressure from LAMMPS & fitted NAG model using PSO method for Al single crystal

### 3.2 Results from Hydrodynamic simulations

We incorporate the NAG model parameters in our one-dimensional metal plate impact hydrodynamic simulations. The plate sizes, target sizes, composition and impact velocities used in our simulations, which correspond to the single crystal experiments [20, 21] are listed in Table-2. For the hydrodynamics simulations of Nb and Cu we used our earlier NAG parameter fits [19]. We get the temporal evolution of the free surface velocity for Mo, Nb, Cu and Al respectively, which are shown in figures 7 to 10.

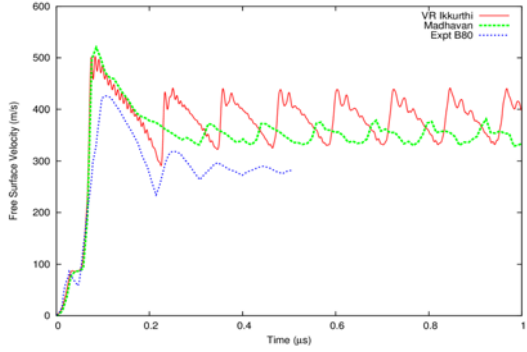

*Fig 7: Free surface velocity vs Time for Mo single crystal (Blue line ref. 20)*

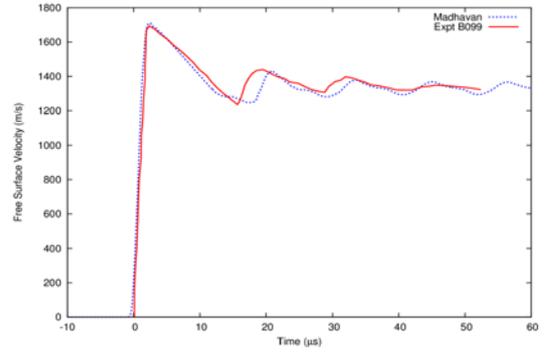

*Fig 8: Free surface velocity vs Time for Nb single crystal (Blue line ref. 20)*

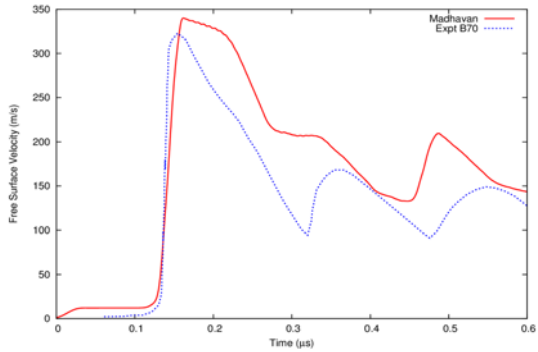

*Fig 9: Free surface velocity vs Time for Cu single crystal (Blue line ref.20)*

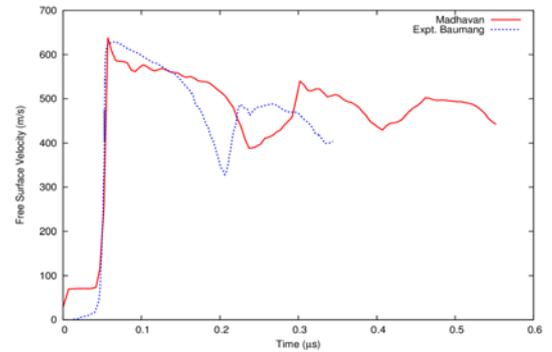

*Fig 10: Free surface velocity vs Time for Al single crystal (Blue line ref. 21)*

Table 2 : Comparison of our simulation results with available experimental results for single crystals of Molybdenum (Mo), Niobium(Nb) & Copper(Cu) [Ref. 20] and Aluminium (Al) [Ref. 21]

| Sl. No. | Material / Expt.Id | Thickness (mm) | | Impact velocity (m/s) | Spall Strength (GPa) | | Spall thickness (mm) | |
|---|---|---|---|---|---|---|---|---|
| | | Flyer | Target | | Simulation | Expt. | Simulation | Expt. |
| 1 | Mo / B80 | 0.40 | 1.38 | 1250 | 5.94 | 5.4 | 0.3 | 0.29 |
| 2 | Mo / B87 | 0.40 | 1.47 | 1250 | 4.83 | 4.6 | 0.294 | 0.29 |
| 3 | Cu / B70 | 0.40 | 4.30 | 660 | 3.96 | 4.5 | 0.332 | 0.33 |
| 4 | Cu / B71 | 0.40 | 4.50 | 660 | 3.71 | 3.95 | 0.301 | 0.3 |
| 5 | Nb / B99 | 0.05 | 0.49 | 4100 | 8.86 | 8.66 | 0.026 | 0.029 |
| 6 | Al/Razornov | 0.40 | 2.90 | 700 | 2.2 | 2.3 | 0.42 | 0.4 |

The deviation of simulations results with experimentally reported results for the spall strength and spall thickness obtained for Mo, Cu and Nb, are less than 10 % in most cases. The multi-scale model is applied to single crystal Al and is reported here for the first time. The values for the NAG model fit for Al are listed in table 1. With these values in our 1D hydrodynamic model, we have simulated for aluminum impact shock experiment results, available from Ref.[21]. Using the free surface velocity as shown in figure 8 and using the formula given for $\sigma_s$ and $t_s$ explained in section 2.3, we obtained the spall strength and spall thickness of single crystal aluminum as 2.2 *GPa* and 0.42 *mm,* compared to the experimental values 2.3 *GPa* and 0.4 *mm,* reported by Razornov et.al.[21]. The good match between the simulations and experiments for Al further validates the multi-scale model.

**4.0 CONCLUSION**

The multi-scale model developed for simulating plate impact has been validated for single crystal Al by obtaining the spall strength and spall thickness and verifying with experiments. The new PSO algorithm and the 1-D hydrodynamic code too have been validated by comparing the results from these with the earlier published results from our multi-scale modeling of Mo, Nb and Cu. Atomistic simulations using the LAMMPS code was used to generate temporal pressure and void volume changes of single crystals subjected to isotropic tensile pressures. Particle Swarm Optimization was used to fit the NAG model parameters to the void volume changes with pressure observed in the MD simulations. The NAG parameters so obtained for single crystals of Mo, Nb, Cu and Al, are used in locally developed 1D hydro-dynamic FORTRAN program to obtain the *spall strength* and *spall thickness*, under impact dynamics. The results are found to match well with the available experimental results.

**ACKNOWLEDGEMENTS**

The authors thank Shri. N Sakthivel and his team for support to use super/parallel computing facility at BARC-Vizag for executing LAMMPS code. The authors also thank their Head of Division, Group Director for their approval to present this work, here.

**REFERENCES**

1. Mark A Meyers, *Dynamic Behavior of Materials*, Wiley-Interscience Pub, (1994)
2. Paul W Cooper, *Explosives Engineering*, Wiley-VCH (1996)
3. Jonas A Zukas and William P Walters, *Explosive Effects and applications,* Springer-Verlag (1998)
4. H. Mark, *Electromagnetic launch technology: The promise and the problems*, IEEE Trans. Magn., **25** (1989) 9-16
5. T. R. Lockner, R. J. Kaye, B. N. Turman Proc. Twenty sixth Intl Symp, *Power Modulator and High Voltage Workshop*, (2004) 119-121.
6. Madhavan S, Sijoy C, Pahari S, Chaturvedi S, *Plasma Sci IEEE Trans* **42** (2014) 323–329.
7. A. K. Gerek, W. R. Thissell, J. N. Johnson, D. L. Tonks and R. Hixson, *J. Mater. Process. Technol.* **60** (1996) 261
8. J.P Fowler, M.J. Worswick, A.K. Pilkey and H.M. Nahme, *Metall. Mater. Trans. A* **31** (2000) 831
9. J. M. Rivas, A. K. Zurek, W. R. Thissell, D. L. Tonks and R. S. Hixson, *Metall. Mater. Trans. A,* **31** (2000) 845
10. R. W. Minich, J. U Cazamias, M Kumar and A. J. Schwartz, *Metall. Mater. Trans. A* **35** (2004) 2663.
11. E. Moshe, S. Eliezer, E. Dekel, A. Ludmirsky, Z. Henis, M. Werdiger, I.B. Goldberg, N. Eliaz and C. Eliezer *J.Appl. Phys.* 83 (1998) 4004
12. T.H. Antoun, L. Seaman, D.R. Curran, *Dynamic Failure of Materials Volume-2-Compilation of Russian Spall Data*, Report DSWA-TR-96-77-V2, Defence Special Weapons, Agency, Alexandria, VA, (1998).
13. S. Plimpton, *J. Comp. Phys.*, **117** (1995) 1-19.
14. D.R. Curran, L. Seaman and D.A. Shockey, *Phys. Rep.,* **147** (1987) 253
15. V.R. Ikkurthi and S. Chaturvedi, *Int. J.Impact Eng.,* **30** (2004) 275-301
16. Mark L Wilkins, *Computer Simulations of Dynamic Phenomena*, Springer-Verlag (1999).
17. S. Rawat, et al, Proc. Of ninth Intl. Conf. *New Models and Hydro-codes for Shock Processes in Condensed Matter*, London, (2012), 27.
18. S. Rawat, V.R. Ikkurthi, M. Warrier, S. Chaturvedi, *PRAMANA Journal of Physics* **83** (2014) 265
19. V. R. Ikkurthi et.al, *Procedia Engineering,* **173** ( 2017 ) 1177 – 1184

20. T.H. Antoun, L. Seaman, D.R. Curran, *Dynamic Failure of Materials Volume-2-Compilation of Russian Spall Data*, Tech. Rep. No DSWA-TR-96-77-V2, Defence Special Weapons, Agency, Alexandria, VA, (1998).

21. S.V Razornov, G.I Kanel, K. Baumung and H. J. Bluhm. *Shock Compression of Condensed Matter* (2001) 503.

22. X. W. Zhou, R. A. Johnson, H. N. G. Wadley, *Phys. Rev. B*, **69** (2004) 1441132014

23. S.M. Foiles, M.I. Baskes and M.S. Daw, *Phys. Rev. B*., **33** (1986) 7983

24. K. W. Jacobsen, J. K. Norskov, and M. J. Puska, *Phys. Rev. B*., **35** (1987) 7423

25. H. Hemani, M. Warrier, S. Chaturvedi, *Journal of Molecular Graphics and Modeling,* **50** (2014) 134-141.

26. Yuhui Shi and Russell Eberhart, Proc. *IEEE international conference on evolutionary computation,* (1998) 69

27. X.-S. Yang, *Test problems in optimization, in: Engineering Optimization: An Introduction with Metaheuristic Applications*, John Wiley and Sons (2010).

28. Sunil Rawat, *Behaviour of solids under high strain rate deformation*, PhD thesis, HBNI, India, (2012)